\documentclass[runningheads]{svmult}
\usepackage{graphicx,physprbb,amsmath,amssymb}
\begin{document}

\title*{Perturbative HQET}
\author{Andrey Grozin}
\institute{Budker Institute of Nuclear Physics, Novosibirsk, Russia}
\maketitle

\begin{abstract}
Methods of calculation of loop diagrams in Heavy Quark Effective Theory (HQET)
are presented.
\end{abstract}

\section{Hadrons with a Heavy Quark}
\label{S:Hadr}

$B$ meson is the hydrogen atom of Quantum Chromodynamics (QCD),
its simplest nontrivial object.
In the leading approximation, the $b$ quark in it
just seats at rest at the origin and creates chromoelectric field.
Light constituents (gluons, light quarks and antiquarks)
move in this external field.
Their motion is relativistic;
the number of gluons and light quark-antiquark pairs in this light cloud
is undetermined and varying.
Therefore, there are no reasons to expect that a nonrelativistic
potential quark model describes $B$ meson well.

Similarly, $\Lambda_b$ baryon can be called the helium atom of QCD.
Unlike in atomic physics, where the hydrogen atom is much
simpler than helium,
$B$ and $\Lambda_b$ are equally difficult.
Both have a light cloud with a variable number of relativistic particles.
The size of this cloud is the confinement radius $1/\Lambda_{\text{QCD}}$;
its properties are determined by the large-distance nonperturbative QCD.

The analogy with atomic physics can tell us a lot about hadrons
with a heavy quark.
The usual hydrogen and tritium have identical chemical properties,
despite the fact that the tritium nucleus is 3 times heavier than the proton.
Both nuclei create identical electric fields, and both stay at rest.
Similarly, $D$ and $B$ mesons have identical ``hadro-chemical'' properties,
despite the fact that $b$ quark is 3 times heavier than $c$.

The proton magnetic moment is of the order of the nuclear magneton $e/(2m_p)$,
and is much smaller than the electron magnetic moment $e/(2m_e)$.
Therefore, the energy difference between the states of the hydrogen atom
with the total spins 0 and 1 (hyperfine splitting) is small
(of the order $m_e/m_p$ times the fine structure).
Similarly, the $b$ quark chromomagnetic moment is proportional to $1/m_b$
by dimensionality, and the hyperfine splitting between $B$ and $B^*$ mesons
is small (proportional to $1/m_b$).

In the limit $m\to\infty$,
the heavy quark spin does not interact with gluon field.
Therefore, it may be rotated at will, without changing physics.
Such rotations can transform $B$ and $B^*$ into each other;
they are degenerate and have identical properties in this limit.
This heavy quark spin symmetry yields many useful relations
among heavy-hadron form factors.
Not only the orientation, but also the magnitude of the heavy quark spin
is irrelevant in the infinite mass limit. 
We can switch off the heavy quark spin, making it spinless,
without affecting physics.

The textbook~\cite{PS:95} contains a good introduction to QCD.
Physics of hadrons with a heavy quark
is discussed in the textbook~\cite{MW:00}.
More examples of perturbative HQET calculations
can be found in the lectures~\cite{G:00a,BG:95a}.

\section{Heavy Quark Effective Theory}
\label{S:HQET}

Let's consider QCD with a heavy flavour $Q$ with mass $m$
and a number of light flavours.
We shall be interested in problems with a single heavy quark
staying approximately at rest.
More exactly, let $\omega\ll m$ be the characteristic momentum scale.
We shall assume that the heavy quark has the momentum $|\vec{p}|\lesssim\omega$
and the energy $|p_0-m|\lesssim\omega$;
light quarks and gluons have momenta $|\vec{k}_i|\lesssim\omega$
and energies $|k_{0i}|\lesssim\omega$.
Heavy quark effective theory (HQET) is an effective field theory
constructed to reproduce QCD results for such problems
expanded up to some order in $\omega/m$.

In our class of problems, the lowest-energy state (``vacuum'')
consists of a single particle -- the heavy quark at rest.
Therefore, it is convenient to use the energy of this state $m$
as a new zero level.
This means that instead of the true energy $p_0$
of the heavy quark (or any state containing this quark)
we shall use the residual energy $\widetilde{p}_0=p_0-m$.
The mass shell of the free heavy quark is
$\widetilde{p}_0=\sqrt{m^2+\vec{p}^2}-m$.
At the leading order in $1/m$, this means that $\widetilde{p}_0=0$
does not depend on $\vec{p}$.
The free Lagrangian giving this dispersion law is
$\widetilde{Q}^+\I D_0\widetilde{Q}$,
where $\widetilde{Q}$ is a 2-component spinor field
describing the heavy quark at rest
(we can also consider it as a 4-component spinor
with the vanishing lower components: $\gamma_0\widetilde{Q}=\widetilde{Q}$).
Reintroducing the interaction with the gluon field
by requirement of the gauge invariance,
we arrive at the HQET Lagrangian
\begin{equation}
L = \widetilde{Q}^+ \I D_0 \widetilde{Q} + \cdots
\label{Lagr}
\end{equation}
where all light-field parts (denoted by dots)
are exactly the same as in QCD.
The field theory~(\ref{Lagr}) is not Lorentz-invariant,
because the heavy quark defines a selected frame -- its rest frame.

The Lagrangian~(\ref{Lagr}) gives the heavy quark propagator
\begin{equation}
\widetilde{S}(\widetilde{p}) = \frac{1}{\widetilde{p}_0+\I0}\,,\quad
\widetilde{S}(x) = -\I\theta(t)\delta(\vec{x})\,.
\label{Prop}
\end{equation}
In the momentum space it depends only on $\widetilde{p}_0$ but not on $\vec{p}$,
because we have neglected the kinetic energy.
Therefore, in the coordinate space the heavy quark does not move.
The unit $2\times2$ matrix is assumed in the propagator~(\ref{Prop}).
It is often convenient to use it as a $4\times4$ matrix;
in such a case, the projector $\frac{1+\gamma_0}{2}$
onto the upper components is assumed.
The static quark interacts only with $A_0$;
the vertex is $\I g v^\mu t^a$,
where $v^\mu=(1,\vec{0})$ is the heavy-quark 4-velocity.
Loops of the heavy quark vanish, because it propagates only forward in time.

The Lagrangian~(\ref{Lagr}) can be rewritten
in covariant notations:
\begin{equation*}
L_v = \overline{\widetilde{Q}}_v \I v\cdot D \widetilde{Q}_v + \cdots
\end{equation*}
where the heavy quark field $\widetilde{Q}_v$ is a 4-component spinor
obeying the relation $\rlap/v\widetilde{Q}_v=\widetilde{Q}_v$,
and $v^\mu$ is the quark 4-velocity.
The momentum $p$ of the heavy quark (or any state containing it)
is related to the residual momentum $\widetilde{p}$ by
\begin{equation*}
p=mv+\widetilde{p}\,, \quad |\widetilde{p}^\mu|\ll m\,.
\end{equation*}
The static quark propagator is
\begin{equation*}
\widetilde{S}(\widetilde{p}) = \frac{1+\rlap/v}{2} \frac{1}{\widetilde{p}\cdot v+\I0}\,,
\end{equation*}
and the vertex is $\I gv^\mu t^a$.

One can watch how expressions for QCD diagrams tend to
the corresponding HQET expressions in the limit $m\to\infty$.
The QCD heavy quark propagator is
\begin{equation*}
S(p) = \frac{\rlap/p+m}{p^2-m^2}
     = \frac{m(1+\rlap/v)+\rlap/\widetilde{p}}{2m\widetilde{p}\cdot v+\widetilde{p}^2}
     = \frac{1+\rlap/v}{2}\frac{1}{\widetilde{p}\cdot v}
       + \mathcal{O}\left(\frac{\widetilde{p}}{m}\right)\,.
\end{equation*}
A vertex $\I g\gamma^\mu t^a$ sandwiched between two projectors
$\frac{1+\rlap/v}{2}$ may be replaced by $\I g v^\mu t^a$.

Renormalization properties (anomalous dimensions etc.) of HQET
differ from those of QCD.
The ultraviolet behavior of an HQET diagram is determined
by the region of loop momenta much larger than the characteristic momentum scale
of the process $\omega$, but much less than the heavy quark mass $m$
(which tends to infinity from the very beginning).
It has nothing to do with the ultraviolet behavior
of the corresponding QCD diagram with the heavy quark line,
which is determined by the region of loop momenta much larger than $m$.

The HQET Lagrangian~(\ref{Lagr})
possesses the $SU(2)$ heavy quark spin symmetry.
If there are $n_{\text{h}}$ heavy-quark fields with the same velocity,
it has the $SU(2n_{\text{h}})$ spin-flavour symmetry.

HQET has great advantages over QCD
in lattice simulation of heavy quark problems.
Indeed, the applicability conditions of the lattice approximation
to problems with light hadrons require
that the lattice spacing $a$ is much less than the characteristic hadron size,
and the total lattice length is much larger than this size.
For simulation of QCD with a heavy quark,
$a$ must be much less than the heavy quark Compton wavelength $1/m$.
For $b$ quark, this is technically impossible at present.
The HQET Lagrangian does not involve the heavy quark mass $m$,
and the applicability conditions of the lattice approximation to HQET
are the same in the case of light hadrons.

\section{One-loop HQET Diagrams}
\label{S:1l}

\begin{figure}[ht]
\begin{center}
\begin{picture}(32,12)
\put(16,6){\makebox(0,0){\includegraphics{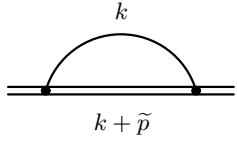}}}
\put(16,-2){\makebox(0,0){$k+\widetilde{p}$}}
\put(16,13){\makebox(0,0){$k$}}
\end{picture}
\end{center}
\caption{One-loop propagator diagram in HQET}
\label{F:1l}
\end{figure}

Now we shall calculate the one-loop HQET propagator diagram
with arbitrary degrees of the denominators (Fig.~\ref{F:1l})
\begin{equation}
I = \int \frac{\D^d k}{(-(k+\widetilde{p})_0-\I0)^{n_1}(-k^2-\I0)^{n_2}}\,.
\label{h1}
\end{equation}
It depends only on $\omega=\widetilde{p}_0$, and not on $\vec{p}$.

The first step is to combine the denominators together.
To this end, we write $1/a^n$ as
\begin{equation*}
\frac{1}{a^n} = \frac{1}{\Gamma(n)}
\int_0^\infty \E^{-a\alpha}\, \alpha^{n-1}\, \D\alpha
\end{equation*}
($\alpha$-representation).
Multiplying two such representations, we have
\begin{equation*}
\frac{1}{a_1^{n_1} a_2^{n_2}} = \frac{1}{\Gamma(n_1) \Gamma(n_2)}
\int \E^{- a_1 \alpha_1 - a_2 \alpha_2}\, \alpha_1^{n_1-1} \alpha_2^{n_2-1}\,
\D\alpha_1\, \D\alpha_2\,.
\end{equation*}
We proceed to the new variables $\alpha_1=y\alpha$, $\alpha_2=\alpha$,
and obtain the HQET Feynman parametrization
\begin{equation}
\frac{1}{a_1^{n_1} a_2^{n_2}} = \frac{\Gamma(n_1+n_2)}{\Gamma(n_1) \Gamma(n_2)}
\int_0^\infty \frac{y^{n_1-1}\, \D y}
{\left[a_1 y + a_2\right]^{n_1+n_2}}\,.
\label{Feyn}
\end{equation}
If the denominator $a_1$ has dimensionality of energy,
and $a_2$ -- of energy squared, then the Feynman parameter $y$
has dimensionality of energy; it runs from 0 to $\infty$.

\begin{figure}[ht]
\begin{center}
\begin{picture}(12,12)
\put(6,6){\makebox(0,0){\includegraphics{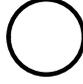}}}
\end{picture}
\end{center}
\caption{Massive one-loop vacuum diagram}
\label{F:vac}
\end{figure}

When combining the denominators in~(\ref{h1}),
it is more convenient to double the linear denominator;
shifting the integration momentum $k\to k-yv$, we obtain
\begin{equation}
I = 2^{n_1} \frac{\Gamma(n_1+n_2)}{\Gamma(n_1) \Gamma(n_2)}
\int \frac{y^{n_1-1}\, \D y\, \D^d k}
{\left[-k^2+y(y-2\omega)-\I0\right]^{n_1+n_2}}\,.
\label{Comb}
\end{equation}
The massive one-loop vacuum diagram (Fig.~\ref{F:vac}) is
\begin{equation*}
\int \frac{\D^d k}{(-k^2+m^2-\I0)^n} =
\I \pi^{d/2} \frac{\Gamma(-d/2+n)}{\Gamma(n)} (m^2)^{d/2-n}\,.
\end{equation*}
Substituting this integral with $m^2\to y(y-2\omega)$ into~(\ref{Comb}),
we obtain at $\omega<0$
\begin{equation*}
I = \I \pi^{d/2} 2^{n_1} \frac{\Gamma(-d/2+n_1+n_2)}{\Gamma(n_1) \Gamma(n_2)}
\int_0^\infty y^{n_1-1}
\left[y(y-2\omega)\right]^{d/2-n_1-n_2}\, \D y\,.
\end{equation*}
We proceed to the dimensionless variable $z=y/(-2\omega)$.
Then the substitution $z+1=1/x$ reduces this integral to the Euler $B$-function:
\begin{equation*}
\begin{split}
&\int_0^\infty y^{d/2-n_2-1} (y-2\omega)^{d/2-n_1-n_2}\, \D y \\
&{} = (-2\omega)^{d-n_1-2n_2}
\frac{\Gamma(-d+n_1+2n_2) \Gamma(d/2-n_2)}{\Gamma(-d/2+n_1+n_2)}\,.
\end{split}
\end{equation*}
Our final result can be written as
\begin{equation}
\begin{split}
& \int \frac{\D^d k}{D_1^{n_1} D_2^{n_2}} =
\I \pi^{d/2} (-2\omega)^{d-2n_2} I(n_1,n_2)\,,\\
& D_1 = (k\cdot v+\omega)/\omega\,, \quad D_2 = -k^2\,,\\
& I(n_1,n_2) =
\frac{\Gamma(-d+n_1+2n_2) \Gamma(d/2-n_2)}{\Gamma(n_1) \Gamma(n_2)}\,.
\end{split}
\label{I1}
\end{equation}

Ultraviolet divergences of this integral are the poles of $\Gamma(-d/2+n_1+2n_2)$,
and infrared ones -- of $\Gamma(d/2-n_2)$.
The integral~(\ref{I1}) has a cut at $\omega>0$,
where the real pair production is possible.
If $n_{1,2}$ are integer, $I(n_1,n_2)$ is proportional to $I_1$, where
\begin{equation}
I_n = \Gamma(1+2n\varepsilon) \Gamma^n(1-\varepsilon)
\label{In}
\end{equation}
is the combination of $\Gamma$ functions which appears
in the $n$-loop HQET sunset diagram.

The one-loop propagator diagram of Fig.~\ref{F:1l}
in the coordinate space is just the product of two propagators.
The HQET propagators in the $p$-space and the $x$-space
are related to each other by the Fourier transform:
\begin{align}
\int_{-\infty}^{+\infty} \frac{\E^{-\I\omega t}}{(-\omega-\I0)^n}
\frac{\D\omega}{2\pi} &{}=
\frac{\I^n}{\Gamma(n)} t^{n-1} \theta(t)\,,
\label{HQET1:Fou1}\\
\int_0^\infty \E^{\I\omega t} t^n\, \D t &{}=
\frac{(-\I)^{n+1} \Gamma(n+1)}{(-\omega-\I0)^{n+1}}\,.
\label{HQET1:Fou2}
\end{align}
For the massless ones,
\begin{align}
\int \frac{\E^{-\I p\cdot x}}{(-p^2-\I0)^n} \frac{\D^d p}{(2\pi)^d} &{}=
\I 2^{-2n} \pi^{-d/2} \frac{\Gamma(d/2-n)}{\Gamma(n)} \frac{1}{(-x^2+\I0)^{d/2-n}}\,,
\label{QCD1:Fou1}\\
\int \frac{\E^{\I p\cdot x}}{(-x^2+\I0)^n} \D^d x &{}=
-\I 2^{d-2n} \pi^{d/2} \frac{\Gamma(d/2-n)}{\Gamma(n)} \frac{1}{(-p^2-i0)^{d/2-n}}
\label{QCD1:Fou2}
\end{align}
Multiplying the HQET propagator~(\ref{HQET1:Fou1}) with the degree $n_1$
and the massless propagator~(\ref{QCD1:Fou1}) with the degree $n_2$, we get
\begin{equation*}
- 2^{-2n_2} \pi^{-d/2}
\frac{\Gamma(d/2-n_2)}{\Gamma(n_1) \Gamma(n_2)}
(\I t)^{n_1+2n_2-d-1} \theta(t)
\end{equation*}
(where $-x^2=(\I t)^2$).
Applying the inverse Fourier transform~(\ref{HQET1:Fou2}),
we reproduce the result~(\ref{I1}).

\section{Renormalization of HQET}
\label{S:Ren}

The HQET Lagrangian expressed via the bare (unrenormalized) quantities
(denoted by the subscript 0) is
\begin{equation*}
\begin{split}
L ={}& \overline{\widetilde{Q}}_0 \I v\cdot D_0 \widetilde{Q}_0
+ \sum_i \bar{q}_{i0} (\I\rlap{\hspace{0.2em}/}D_0-m_{i0}) q_{i0}
- \frac{1}{4} G^a_{0\mu\nu} G^{a\mu\nu}_0\\
&{} - \frac{1}{2a_0} \left(\partial_\mu A_0^\mu\right)^2
+ (\partial_\mu \bar{c}_0^a) (D_0^\mu c_0^a)\,,
\end{split}
\end{equation*}
where $D_{0\mu}q_0=(\partial_\mu-i g_0 A_{0\mu}^a t^a)q_0$,
$a_0$ is the gauge fixing parameter, and $c$ is the ghost field.
The renormalized quantities are related to the bare ones by
\begin{equation}
\begin{split}
&\widetilde{Q}_0 = \widetilde{Z}_Q^{1/2} \widetilde{Q}\,, \quad
q_{i0} = Z_q^{1/2} q_i\,, \quad
A_0 = Z_A^{1/2} A\,, \quad
c_0 = Z_c^{1/2} c\,, \\
&g_0 = Z_\alpha^{1/2} g\,, \quad
m_{i0} = Z_m m_i\,, \quad
a_0 = Z_A a\,,
\end{split}
\label{renorm}
\end{equation}
where renormalization factors have the minimal structure
\begin{equation}
Z = 1 + \frac{Z_{11}}{\varepsilon} \frac{\alpha_{\text{s}}}{4\pi}
+ \left(\frac{Z_{22}}{\varepsilon^2}+\frac{Z_{21}}{\varepsilon}\right)
\left(\frac{\alpha_{\text{s}}}{4\pi}\right)^2 + \cdots
\label{minim}
\end{equation}
All the renormalization constants in~(\ref{renorm})
are the same as in QCD,
where the heavy flavour $Q$ is not counted in $n_{\text{f}}$.
In order to find the new constant $\widetilde{Z}_Q$,
we need to calculate the heavy quark propagator in HQET.

The Lagrangian has dimensionality $[L]=d$,
because the action $S = \int L \D^d x$
is exactly dimensionless in the space-time with any $d$.
The gluon kinetic term has the structure $(\partial A_0)^2$;
hence, the dimensionality of the gluon field is $[A_0]=1-\varepsilon$.
Similarly, from the quark kinetic terms, the dimensionality
of the quark fields is $[q_{i0}]=\frac{3}{2}-\varepsilon$.
The covariant derivative $D_{0\mu}=\partial_\mu-i g_0 A_{0\mu}^a t^a$
has dimensionality 1, hence the dimensionality of the coupling constant
is $[g_0]=\varepsilon$.

We define $\alpha_{\text{s}}$ to be exactly dimensionless:
\begin{equation}
\frac{\alpha_{\text{s}}(\mu)}{4\pi} = \mu^{-2\varepsilon} \frac{g^2}{(4\pi)^{d/2}}
\E^{-\gamma\varepsilon}
\quad\text{or}\quad
\frac{g_0^2}{(4\pi)^{d/2}} = \mu^{2\varepsilon} \frac{\alpha_{\text{s}}(\mu)}{4\pi}
Z_\alpha(\alpha_{\text{s}}(\mu)) \E^{\gamma\varepsilon}\,.
\label{alphas}
\end{equation}
Here $\mu$ is the renormalization scale,
and the factor $\exp(-\gamma\varepsilon)$ is included for convenience.
All the bare quantities, including $g_0$, are $\mu$-independent.
Differentiating~(\ref{alphas}),
we obtain the renormalization group equation for $\alpha_{\text{s}}(\mu)$:
\begin{equation}
\begin{split}
\frac{\D\log\alpha_{\text{s}}}{\D\log\mu} &= -2\varepsilon-2\beta(\alpha_{\text{s}})\,, \\
\beta(\alpha_{\text{s}}) &= \frac{1}{2} \frac{\D\log Z_\alpha}{\D\log\mu}
= \beta_0 \frac{\alpha_{\text{s}}}{4\pi}
+ \beta_1 \left(\frac{\alpha_{\text{s}}}{4\pi}\right)^2 + \cdots
\end{split}
\label{rengroup}
\end{equation}
where
\begin{equation*}
\beta_0 = \frac{11}{3} C_{\text{A}} - \frac{4}{3} T_{\text{F}} n_{\text{f}}\,,\ldots
\end{equation*}

\begin{figure}[ht]
\begin{center}
\begin{picture}(111,9)
\put(55.5,4.5){\makebox(0,0){\includegraphics{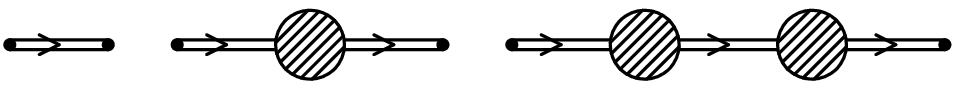}}}
\put(14.5,4.5){\makebox(0,0){+}}
\put(48.5,4.5){\makebox(0,0){+}}
\put(99.5,4.5){\makebox(0,0){+}}
\put(103,4.5){\makebox(0,0)[l]{$\cdots$}}
\end{picture}
\end{center}
\caption{Structure of diagrams for the heavy-quark propagator in HQET}
\label{F:struct}
\end{figure}

The bare (unrenormalized) heavy quark propagator $\I\widetilde{S}(\omega)$
has the structure (Fig.~\ref{F:struct})
\begin{equation*}
\begin{split}
\I\widetilde{S}(\omega) ={}& \I\widetilde{S}_0(\omega)
+ \I\widetilde{S}_0(\omega) (-\I)\widetilde{\Sigma}(\omega) \I\widetilde{S}_0(\omega)\\
&{} + \I\widetilde{S}_0(\omega) (-\I)\widetilde{\Sigma}(\omega) \I\widetilde{S}_0(\omega)
(-\I)\widetilde{\Sigma}(\omega) \I\widetilde{S}_0(\omega) + \cdots
\end{split}
\end{equation*}
where $\widetilde{S}_0(\omega) = 1/\omega$ is the free HQET propagator,
and the heavy-quark self-energy $-\I\widetilde{\Sigma}(\omega)$
is the sum of one-particle-irreducible HQET self-energy diagrams
(which cannot be separated into two disconnected parts
by cutting a single heavy quark line).
Summing this series, we obtain
\begin{equation*}
\widetilde{S}(\omega) = \frac{1}{\omega-\widetilde{\Sigma}(\omega)}\,.
\end{equation*}

\begin{figure}[ht]
\begin{center}
\begin{picture}(32,9.5)
\put(16,4.75){\makebox(0,0){\includegraphics{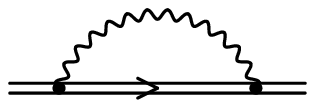}}}
\end{picture}
\end{center}
\caption{One-loop heavy-quark self-energy in HQET}
\label{F:S1}
\end{figure}

In the one-loop approximation (Fig.~\ref{F:S1}),
\begin{equation*}
\widetilde{\Sigma}(\omega) = \I C_{\text{F}} \int \frac{\D^d k}{(2\pi)^d}
\I g_0 v^\mu \frac{\I}{k\cdot v+\omega} \I g_0 v^\nu \frac{-\I}{k^2}
\left( g_{\mu\nu} - (1-a_0) \frac{k_\mu k_\nu}{k^2} \right)\,.
\end{equation*}
After contraction over the indices,
the second term in the brackets contains
$(k\cdot v)^2=(k\cdot v+\omega-\omega)^2$.
This factor can be replaced by $\omega^2$, because all integrals
without $k\cdot v+\omega$ in the denominator are scale-free and hence vanish.
Using the definition~(\ref{I1}), we get
\begin{equation*}
\widetilde{\Sigma}(\omega) = C_{\text{F}}
\frac{g_0^2(-2\omega)^{1-2\varepsilon}}{(4\pi)^{d/2}}
\left[ 2 I(1,1) + \tfrac{1}{2} (1-a_0) I(1,2) \right]\,,
\end{equation*}
and, finally,
\begin{equation}
\widetilde{\Sigma}(\omega) = - C_{\text{F}}
\frac{g_0^2(-2\omega)^{1-2\varepsilon}}{(4\pi)^{d/2}}
\frac{I_1}{d-4} \left(a_0-1-\frac{2}{d-3}\right)\,,
\label{Si1}
\end{equation}
where $I_1$ is defined in~(\ref{In}).

Therefore, with the one-loop accuracy, the heavy quark propagator in HQET is
\begin{equation*}
\omega \widetilde{S}(\omega) =
1 + C_{\text{F}} \frac{g_0^2(-2\omega)^{-2\varepsilon}}{(4\pi)^{d/2}}
\frac{I_1}{d-4} 2 \left(a_0-1-\frac{2}{d-3}\right)\,.
\end{equation*}
\index{coordinate space|(}
Now we re-express it via the renormalized quantities $\alpha_{\text{s}}(\mu)$, $a(\mu)$.
Expanding the result in $\varepsilon$ up to the finite term, we get
\begin{equation*}
\omega \widetilde{S}(\omega) = 1 +
C_{\text{F}} \frac{\alpha_{\text{s}}(\mu)}{4\pi\varepsilon} \E^{-2L\varepsilon}
\left(3 - a(\mu) + 4\varepsilon\right)
\quad\text{where}\quad
L = \log\frac{-2\omega}{\mu}\,.
\end{equation*}
This should be equal to
$\widetilde{Z}_Q(\alpha_{\text{s}}(\mu),a(\mu)) \omega \widetilde{S}_{\text{r}}(\omega;\mu)$,
where the renormalization constant $\widetilde{Z}_Q(\alpha_{\text{s}},a)$
has the minimal form~(\ref{minim}),
and the renormalized propagator $\widetilde{S}_{\text{r}}(\omega;\mu)$
is finite at $\varepsilon\to0$.
We obtain
\begin{align}
\widetilde{Z}_Q(\alpha_{\text{s}},a) &{}= 1 + C_{\text{F}}
\frac{\alpha_{\text{s}}}{4\pi\varepsilon} (3-a)\,,
\label{ZQ1}\\
\omega \widetilde{S}_{\text{r}}(\omega;\mu) &{}= 1 + C_{\text{F}}
\frac{\alpha_{\text{s}}(\mu)}{4\pi} 2
\left[ (a(\mu)-3) L + 2 \right]\,.
\nonumber
\end{align}

For any renormalization constant~(\ref{minim}),
the corresponding anomalous dimension is defined by
\begin{equation*}
\gamma(\alpha_{\text{s}}) = \frac{\D\log Z}{\D\log\mu} =
\gamma_0 \frac{\alpha_{\text{s}}}{4\pi}
+ \gamma_1 \left(\frac{\alpha_{\text{s}}}{4\pi}\right)^2 + \cdots
\end{equation*}
It is more convenient to present results for anomalous dimensions
instead of renormalization constants, because anomalous dimensions
contain the same information but are more compact.
With the one-loop accuracy, it is sufficient to use
the zeroth-order term $-2\varepsilon$ in~(\ref{rengroup})
when differentiating $\alpha_{\text{s}}(\mu)$.
We obtain
\begin{equation*}
\widetilde{\gamma}_Q = 2 C_{\text{F}} \frac{\alpha_{\text{s}}}{4\pi} (a-3) + \cdots
\end{equation*}
Note this anomalous dimension vanishes in the Yennie gauge $a=3$.

\section{Two-loop HQET Diagrams}
\label{S:2l}

\begin{figure}[ht]
\begin{center}
\begin{picture}(112,47)
\put(56,26){\makebox(0,0){\includegraphics{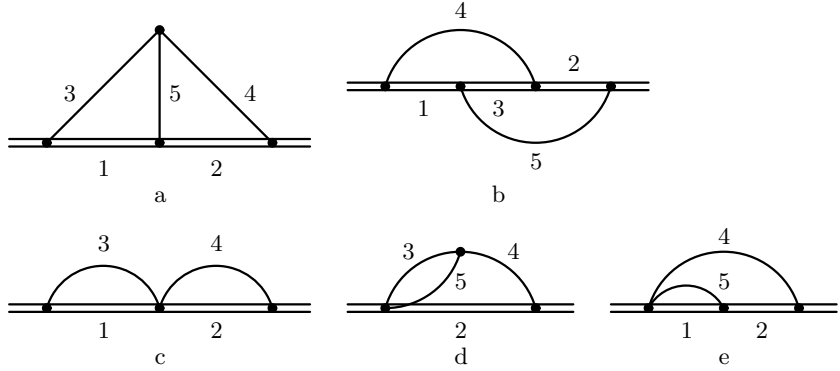}}}
\put(13.5,26){\makebox(0,0){1}}
\put(28.5,26){\makebox(0,0){2}}
\put(9,36){\makebox(0,0){3}}
\put(33,36){\makebox(0,0){4}}
\put(23,36){\makebox(0,0){5}}
\put(21,21.5){\makebox(0,0)[b]{a}}
\put(56,34){\makebox(0,0){1}}
\put(66,34){\makebox(0,0){3}}
\put(76,40){\makebox(0,0){2}}
\put(61,47){\makebox(0,0){4}}
\put(71,27){\makebox(0,0){5}}
\put(66,21.5){\makebox(0,0)[b]{b}}
\put(13.5,4.5){\makebox(0,0){1}}
\put(28.5,4.5){\makebox(0,0){2}}
\put(13.5,16){\makebox(0,0){3}}
\put(28.5,16){\makebox(0,0){4}}
\put(21,0){\makebox(0,0)[b]{c}}
\put(61,4.5){\makebox(0,0){2}}
\put(54,15){\makebox(0,0){3}}
\put(68,15){\makebox(0,0){4}}
\put(61,11){\makebox(0,0){5}}
\put(61,0){\makebox(0,0)[b]{d}}
\put(91,4.5){\makebox(0,0){1}}
\put(101,4.5){\makebox(0,0){2}}
\put(96,17){\makebox(0,0){4}}
\put(96,11){\makebox(0,0){5}}
\put(96,0){\makebox(0,0)[b]{e}}
\end{picture}
\end{center}
\caption{Two-loop HQET propagator diagram}
\label{F:2l}
\end{figure}

There are two generic topologies
of two-loop HQET propagator diagrams, Fig.~\ref{F:2l}a, b.
If one of the lines is shrunk into a point,
the diagrams of Fig.~\ref{F:2l}c, d, e result.
If any two adjacent lines are shrunk into a point,
the diagram contains a no-scale vacuum tadpole,
and hence vanishes.

We write down the diagram of Fig.~\ref{F:2l}a as
\begin{align}
&\int \frac{\D^d k_1\, \D^d k_2}{D_1^{n_1} D_2^{n_2} D_3^{n_3} D_4^{n_4} D_5^{n_5}}
= - \pi^d (-2\omega)^{2(d-n_3-n_4-n_5)} I(n_1,n_2,n_3,n_4,n_5)\,,
\nonumber\\
&D_1 = (k_1+\widetilde{p})\cdot v/\omega\,,\quad
D_2 = (k_2+\widetilde{p})\cdot v/\omega\,,
\label{h2}\\
&D_3 = -k_1^2\,,\quad D_4 = -k_2^2\,,\quad D_5 = -(k_1-k_2)^2\,.
\nonumber
\end{align}
It is symmetric with respect to $(1\leftrightarrow3,2\leftrightarrow4)$.
If indices of any two adjacent lines are zero,
the diagram contains a scale-free vacuum subdiagram,
and hence vanishes.

\begin{figure}[ht]
\begin{center}
\begin{picture}(102,27)
\put(51,13.5){\makebox(0,0){\includegraphics{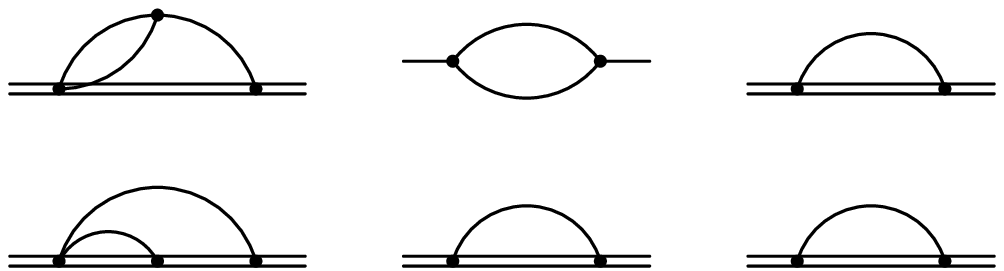}}}
\put(16,16.5){\makebox(0,0)[t]{$n_2$}}
\put(9,25){\makebox(0,0)[b]{$n_3$}}
\put(23,25){\makebox(0,0)[b]{$n_4$}}
\put(17,21.5){\makebox(0,0){$n_5$}}
\put(36,21.3125){\makebox(0,0){$=$}}
\put(53.5,15.5625){\makebox(0,0)[t]{$n_5$}}
\put(53.5,26.0625){\makebox(0,0)[b]{$n_3$}}
\put(71,21.3125){\makebox(0,0){$\times$}}
\put(88.5,16.5){\makebox(0,0)[t]{$n_2$}}
\put(88.5,25.125){\makebox(0,0)[b]{$n_4+n_3+n_5-d/2$}}
\put(11,-1){\makebox(0,0)[t]{$n_1$}}
\put(21,-1){\makebox(0,0)[t]{$n_2$}}
\put(16,9.5){\makebox(0,0)[b]{$n_4$}}
\put(17,4.5){\makebox(0,0){$n_5$}}
\put(36,3.8125){\makebox(0,0){$=$}}
\put(53.5,-1){\makebox(0,0)[t]{$n_1$}}
\put(53.5,7.625){\makebox(0,0)[b]{$n_5$}}
\put(71,3.8125){\makebox(0,0){$\times$}}
\put(88.5,-1){\makebox(0,0)[t]{$n_2+n_1+2n_5-d$}}
\put(88.5,7.625){\makebox(0,0)[b]{$n_4$}}
\end{picture}
\end{center}
\caption{Trivial two-loop diagrams}
\label{F:triv}
\end{figure}

\begin{figure}[ht]
\begin{center}
\begin{picture}(32,24)
\put(16,12){\makebox(0,0){\includegraphics{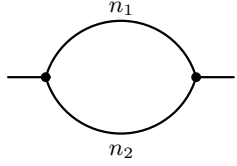}}}
\put(16,2){\makebox(0,0){$n_2$}}
\put(16,21){\makebox(0,0){$n_1$}}
\end{picture}
\end{center}
\caption{Massless one-loop propagator diagram}
\label{F:m0}
\end{figure}

When one of the indices is zero, this diagram is trivial (Fig.~\ref{F:triv}).
If $n_5=0$, this is a product of two one-loop diagrams (Fig.~\ref{F:2l}c):
\begin{equation}
I(n_1,n_2,n_3,n_4,0) = I(n_1,n_3) I(n_2,n_4)\,.
\label{I50}
\end{equation}
If $n_1=0$ (Fig.~\ref{F:2l}d), the $(3,5)$ integral
gives $G(n_3,n_5)/(-k_2^2)^{n_3+n_5-d/2}$;
this is combined with the denominator 4, and we obtain
\begin{equation}
I(0,n_2,n_3,n_4,n_5) = G(n_3,n_5) I(n_2,n_4+n_3+n_5-d/2)
\label{I10}
\end{equation}
(and similarly for $n_2=0$).
If $n_3=0$ (Fig.~\ref{F:2l}e), the $(1,5)$ integral~(\ref{I1})
gives $I(n_1,n_5)\allowbreak/(-2\omega)^{n_1+2n_5-d}$;
this is combined with the denominator 2, and we obtain
\begin{equation}
I(n_1,n_2,0,n_4,n_5) = I(n_1,n_5) I(n_2+n_1+2n_5-d,n_4)
\label{I30}
\end{equation}
(and similarly for $n_4=0$).
Here $G(n_1,n_2)$ is the massless one-loop propagator integral (Fig.~\ref{F:m0}):
\begin{equation*}
\begin{split}
& \int \frac{\D^d k}{D_1^{n_1} D_2^{n_2}} =
\I \pi^{d/2} (-p^2)^{d/2-n_1-n_2} G(n_1,n_2)\,,\\
& D_1 = -(k+p)^2\,, \quad D_2 = -k^2\,,\\
& G(n_1,n_2) = \frac{\Gamma(-d/2+n_1+n_2) \Gamma(d/2-n_1) \Gamma(d/2-n_2)}
{\Gamma(n_1) \Gamma(n_2) \Gamma(d-n_1-n_2)}\,,
\end{split}
\end{equation*}

When all $n_i>0$, the problem does not immediately reduce
to a repeated use of the one-loop formulae.
We shall use a powerful method
called integration by parts~\cite{CT:81}.
It is based on the simple observation that any integral of
$\partial/\partial k_1(\cdots)$
(or $\partial/\partial k_2(\cdots)$) vanishes.
From this, we can obtain recurrence relations which involve
$I(n_1,n_2,n_3,n_4,n_5)$ with different sets of indices.
Applying these relations in a carefully chosen order,
we can reduce any $I(n_1,n_2,n_3,n_4,n_5)$ to trivial ones.

The differential operator $\partial/\partial k_2$
applied to the integrand of~(\ref{h2}) acts as
\begin{equation}
\frac{\partial}{\partial k_2} \to
-\frac{n_2}{D_2}\frac{v}{\omega} + \frac{n_4}{D_4}2k_2 + \frac{n_5}{D_5}2(k_2-k_1)\,.
\label{HQET2:dk1}
\end{equation}
Applying $(\partial/\partial k_2)\cdot(k_2-k_1)$ to the integrand of~(\ref{h2}),
we get a vanishing integral.
On the other hand, from~(\ref{HQET2:dk1}), $(k_2-k_1)\cdot v/\omega=D_2-D_1$,
$2(k_2-k_1)\cdot k_2=D_3-D_4-D_5$, we see that this differential operator
is equivalent to inserting
\begin{equation*}
d-n_2-n_4-2n_5 + \frac{n_2}{D_2}D_1 + \frac{n_4}{D_4}(D_3-D_5)
\end{equation*}
under the integral sign
($d$ comes from differentiating $(k_2-k_1)$).
Therefore, we obtain the recurrence relation~\cite{BG:91}
\begin{equation}
\left[ d-n_2-n_4-2n_5 + n_2\mathbf2^+\mathbf1^- + n_4\mathbf4^+(\mathbf3^--\mathbf5^-) \right] I = 0\,,
\label{Tri2}
\end{equation}
where, for example,
\begin{equation*}
\mathbf1^\pm I(n_1,n_2,n_3,n_4,n_5) = I(n_1\pm1,n_2,n_3,n_4,n_5)\,.
\end{equation*}

Expressing $I(n_1,n_2,n_3,n_4,n_5)$ from~(\ref{Tri2}):
\begin{equation}
I(n_1,n_2,n_3,n_4,n_5) =
\frac{n_4\mathbf4^+(\mathbf5^--\mathbf3^-) - n_2\mathbf2^+\mathbf1^-}{d-n_2-n_4-2n_5} I\,,
\label{Tri}
\end{equation}
we see that the sum $n_1+n_3+n_5$ reduces by 1.
Therefore, applying~(\ref{Tri}) sufficiently many times,
we can reduce an arbitrary $I$ integral with integer indices
to a combination of integrals
with $n_5=0$ (Fig.~\ref{F:2l}c, (\ref{I50})),
$n_1=0$ (Fig.~\ref{F:2l}d, (\ref{I10})),
$n_3=0$ (Fig.~\ref{F:2l}e, (\ref{I30})).
Thus, any integral $I(n_1,n_2,n_3,n_4,n_5)$ with integer $n_i$
can be expressed as a linear combination of $I_1^2$ and $I_2$ (\ref{In}),
coefficients being rational functions of $d$.

The second two-loop topology, Fig.~\ref{F:2l}b,
\begin{align}
&\int \frac{\D^d k_1\, \D^d k_2}{D_1^{n_1} D_2^{n_2} D_3^{n_3} D_4^{n_4} D_5^{n_5}}
= - \pi^d (-2\omega)^{2(d-n_4-n_5)} J(n_1,n_2,n_3,n_4,n_5)\,,
\nonumber\\
&D_1 = (k_1+\widetilde{p})\cdot v/\omega\,,\quad
D_2 = (k_2+\widetilde{p})\cdot v/\omega\,,\quad
D_3 = (k_1+k_2+\widetilde{p})\cdot v/\omega\,,
\nonumber\\
&D_4 = -k_1^2\,,\quad D_5 = -k_2^2\,,
\label{j2}
\end{align}
is much easier.
If $n_4=0$, or $n_5=0$, or two adjacent heavy indices are zero,
the diagram vanishes.
If $n_3=0$, this is a product of two one-loop diagrams (Fig.~\ref{F:2l}c):
\begin{equation*}
J(n_1,n_2,0,n_4,n_5) = I(n_1,n_4) I(n_2,n_5)\,.
\end{equation*}
If $n_1=0$ this is a diagram of Fig.~\ref{F:2l}e:
\begin{equation*}
J(0,n_2,n_3,n_4,n_5) = I(n_3,n_4) I(n_2+n_3+2n_4-d,n_5)\,;
\end{equation*}
the case $n_2=0$ is symmetric.
The heavy denominators are linearly dependent:
\begin{equation*}
1 = D_1 + D_2 - D_3
\end{equation*}
When $n_{1,2,3}$ are all positive, we can insert this
into the integrand of~(\ref{j2}), and obtain~\cite{BG:91}
\begin{equation}
J = (\mathbf1^-+\mathbf2^--\mathbf3^-) J\,.
\label{parfrac}
\end{equation}
This reduces $n_1+n_2+n_3$ by 1.
Therefore, applying~(\ref{parfrac}) sufficiently many times,
we can reduce an arbitrary $J$ integral with integer indices
to a combination of trivial integrals with $n_1=0$, $n_2=0$, $n_3=0$.

\begin{figure}[ht]
\begin{center}
\begin{picture}(69,34.5)
\put(34.5,17.25){\makebox(0,0){\includegraphics{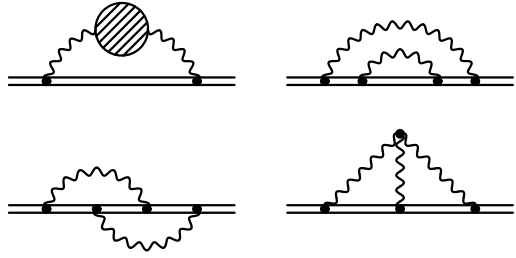}}}
\end{picture}
\end{center}
\caption{Two-loop heavy-quark self-energy in HQET}
\label{F:S2}
\end{figure}

Using these methods, it is easy to calculate the two-loop heavy-quark self-energy
(Fig.~\ref{F:S2}).
From this result, the two-loop renormalization of the heavy-quark field in HQET
can be obtained~\cite{BG:91}:
\begin{equation}
\widetilde{\gamma}_Q = 2 (a-3) C_{\text{F}} \frac{\alpha_{\text{s}}}{4\pi}
+ C_{\text{F}} \left( \frac{3a^2+24a-179}{6} C_{\text{A}} + \frac{32}{3} T_{\text{F}} n_{\text{f}} \right)
\left(\frac{\alpha_{\text{s}}}{4\pi}\right)^2 + \cdots
\label{gammaQ2}
\end{equation}

The method of calculation of three-loop propagator diagrams in HQET,
based on integration by parts, has been constructed in~\cite{G:00}

\section{Heavy Electron Effective Theory}
\label{S:HEET}

Now we make a short digression into the abelian version of HQET -- the
Heavy Electron Effective Theory, an effective field theory of QED
describing interaction of a single electron with soft photons.
It is obtained by setting $C_{\text{F}}\to1$, $C_{\text{A}}\to0$,
$g_0\to e_0$, $\alpha_{\text{s}}\to\alpha$.
This theory was considered long ago,
and is called the Bloch--Nordsieck model.

\begin{figure}[ht]
\begin{center}
\begin{picture}(110.6,42.7)
\put(55.3,21.35){\makebox(0,0){\includegraphics[scale=0.7]{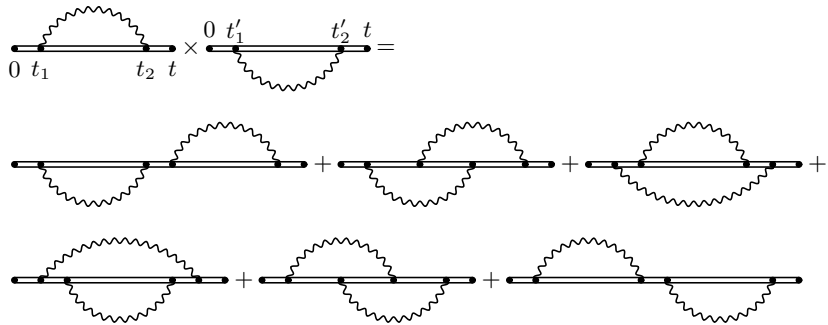}}}
\put(0.7,35.0){\makebox(0,0)[t]{0}}
\put(4.2,35.0){\makebox(0,0)[t]{$t_1$}}
\put(18.2,35.0){\makebox(0,0)[t]{$t_2$}}
\put(21.7,35.0){\makebox(0,0)[t]{$t$}}
\put(26.6,40.25){\makebox(0,0)[t]{0}}
\put(30.1,40.6){\makebox(0,0)[t]{$t_1'$}}
\put(44.1,40.6){\makebox(0,0)[t]{$t_2'$}}
\put(47.6,40.25){\makebox(0,0)[t]{$t$}}
\put(24.15,36.75){\makebox(0,0){$\times$}}
\put(50.05,36.75){\makebox(0,0){=}}
\put(41.65,21.35){\makebox(0,0){+}}
\put(74.55,21.35){\makebox(0,0){+}}
\put(107.45,21.35){\makebox(0,0){+}}
\put(31.15,5.95){\makebox(0,0){+}}
\put(64.05,5.95){\makebox(0,0){+}}
\end{picture}
\end{center}
\caption{Exponentiation theorem}
\label{F:exp}
\end{figure}

Suppose we calculate the one-loop correction to the heavy electron propagator
in the coordinate space.
Let's multiply this correction by itself (Fig.~\ref{F:exp}).
We get an integral in $t_1$, $t_2$, $t_1'$, $t_2'$
with $0<t_1<t_2<t$, $0<t_1'<t_2'<t$.
Ordering of primed and non-primed integration times can be arbitrary.
The integration area is subdivided into 6 regions,
corresponding to 6 diagrams in Fig.~\ref{F:exp}.
This is twice the two-loop correction to the propagator.
Continuing this drawing exercise,
we see that the one-loop correction cubed
is $3!$ times the three-loop correction, and so on.
Therefore, the exact all-order propagator
is the exponent of the one-loop term:
\begin{equation}
\widetilde{S}(t) = \widetilde{S}_0(t) \exp \left[ \frac{e_0^2 (\I t/2)^{2\varepsilon}}{(4\pi)^{d/2}}
\Gamma(-\varepsilon) \left(a_0 - 1 - \frac{2}{d-3}\right) \right]\,.
\label{HEET:S}
\end{equation}

\begin{figure}[ht]
\begin{center}
\begin{picture}(86,12.5)
\put(43,7.75){\makebox(0,0){\includegraphics{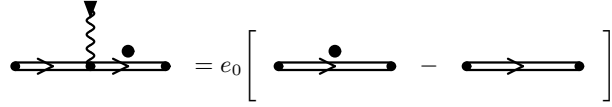}}}
\put(28.5,4){\makebox(0,0){${}=e_0\Biggl[\Biggr.$}}
\put(56,4){\makebox(0,0){$-$}}
\put(80,4){\makebox(0,0){$\Biggl.\Biggr]$}}
\end{picture}
\end{center}
\caption{Ward identity for the free electron propagator}
\label{F:WardS}
\end{figure}

In this theory, $Z_A=1$, because there exist no loops
which can be inserted into the photon propagator.
Now we are going to show that $Z_\alpha=1$, too.
To this end, let's consider the sum of all one-particle-irreducible
vertex diagrams, not including the external leg propagators --
the electron--photon proper vertex.
It has the same structure as the tree-level term:
$\I e_0 v^\mu \widetilde{\Gamma}$, $\widetilde{\Gamma}=1+\widetilde{\Lambda}$,
where $\widetilde{\Lambda}$ is the sum of all unrenormalized diagrams
starting from one loop.
Let's multiply the vertex by the incoming photon momentum $q_\mu$.
This product can be simplified by the Ward identity
for the electron propagator (Fig.~\ref{F:WardS}):
\begin{equation*}
\begin{split}
\I \widetilde{S}_0(\widetilde{p}') \; \I e_0 v\cdot q \; \I \widetilde{S}_0(\widetilde{p}) &{}=
\I e_0 \frac{\I}{\widetilde{p}'\cdot v} \left[ \widetilde{p}'\cdot v - \widetilde{p}\cdot v \right]
\frac{\I}{\widetilde{p}\cdot v}\\
&{}= \I e_0 \left[ \widetilde{S}_0(\widetilde{p}') - \widetilde{S}_0(\widetilde{p}) \right]\,.
\end{split}
\end{equation*}
In the Figure, the gluon line with the black triangle at the end
means that the vertex is contracted with the incoming gluon momentum $q$
(it includes no gluon propagator!);
a dot near an electron propagator means that its momentum
is shifted by $q$.

\begin{figure}[ht]
\begin{center}
\begin{picture}(106,140)
\put(53,70){\makebox(0,0){\includegraphics{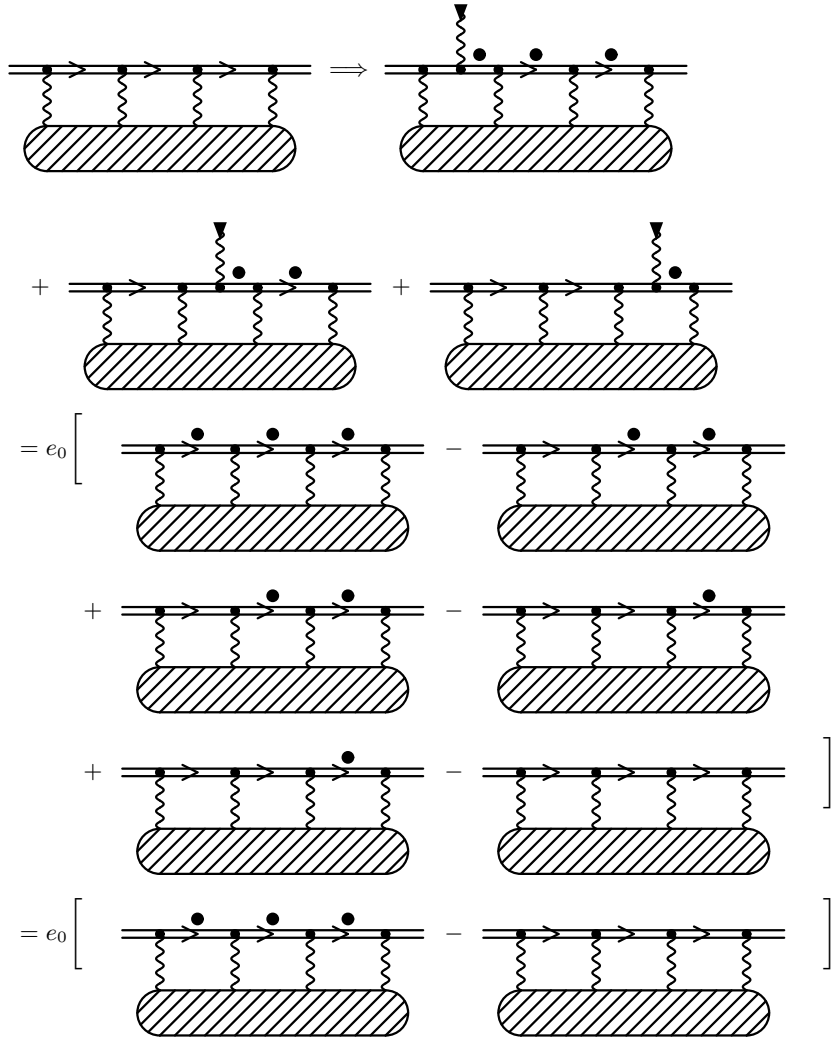}}}
\put(46,129.25){\makebox(0,0){$\Longrightarrow$}}
\put(5,100.5){\makebox(0,0){$+$}}
\put(53,100.5){\makebox(0,0){$+$}}
\put(6,79){\makebox(0,0){${}=e_0\Biggl[\Biggr.$}}
\put(60,79){\makebox(0,0){$-$}}
\put(12,57.5){\makebox(0,0){$+$}}
\put(60,57.5){\makebox(0,0){$-$}}
\put(12,36){\makebox(0,0){$+$}}
\put(60,36){\makebox(0,0){$-$}}
\put(110,36){\makebox(0,0){$\Biggl.\Biggr]$}}
\put(6,14.5){\makebox(0,0){${}=e_0\Biggl[\Biggr.$}}
\put(60,14.5){\makebox(0,0){$-$}}
\put(110,14.5){\makebox(0,0){$\Biggl.\Biggr]$}}
\end{picture}
\end{center}
\caption{Ward identity for the electron--photon vertex}
\label{F:WardV}
\end{figure}

Starting from each diagram for $\widetilde{\Sigma}$,
we can obtain a set of diagrams for $\widetilde{\Lambda}$
by inserting the external photon vertex into each electron propagator.
After multiplying by $q_\mu$, each diagram in this set becomes a difference.
All terms cancel each other, except the extreme ones (Fig.~\ref{F:WardV}),
and we obtain the Ward identity
\begin{equation}
\widetilde{\Lambda}(\omega,\omega') =
- \frac{\widetilde{\Sigma}(\omega')-\widetilde{\Sigma}(\omega)}{\omega'-\omega}
\quad\text{or}\quad
\widetilde{\Gamma}(\omega',\omega) =
\frac{\widetilde{S}^{-1}(\omega')-\widetilde{S}^{-1}(\omega)}{\omega'-\omega}\,.
\label{wardV}
\end{equation}
The vertex function is thus also known to all orders.
The charge renormalization constant $Z_\alpha$ is obtained from the requirement
that the renormalized vertex function $g_0\widetilde{\Gamma}Z_A^{1/2}\widetilde{Z}_Q$ is finite.
The factor $\widetilde{Z}_Q$ transforms $\widetilde{S}^{-1}$ in~(\ref{wardV})
into $\widetilde{S}_{\text{r}}^{-1}$ and hence makes $\widetilde{\Gamma}$ finite.
Therefore, the remaining factor $(Z_\alpha Z_A)^{1/2}=1$
(this is also true in QED).
In the Bloch--Nordsieck model, $Z_A=1$ and hence $Z_\alpha=1$.

Due to the absence of charge and photon field renormalization,
we may replace $e_0\to e$, $a_0\to a$ in the bare propagator~(\ref{HEET:S}).
It is made finite by the minimal (in the sense of~(\ref{minim}))
renormalization constant, which is just the exponent of the one-loop term
\begin{equation*}
\widetilde{Z}_Q = \exp \left[ -(a-3)\frac{\alpha}{4\pi\varepsilon} \right]\,,
\end{equation*}
and the anomalous dimension is exactly equal to the one-loop contribution
\begin{equation*}
\widetilde{\gamma}_Q = 2 (a-3) \frac{\alpha}{4\pi}\,.
\end{equation*}
Note that the electron propagator is finite to all orders in the Yennie gauge.

What information useful for the real HQET can be extracted from this
simple abelian model?
We can obtain the $C_{\text{F}}^2$ term in the two-loop propagator $\widetilde{S}(\omega)$
without explicit calculation.
There should be no $C_{\text{F}}^2$ term
in the two-loop field anomalous dimension~(\ref{gammaQ2}).

\section{Heavy-light Quark Currents}
\label{S:hl}

Heavy-light quark currents are often useful
(e.g., the weak $b\to u$ current).
QCD currents are expanded in $1/m$;
coefficients of these expansions contain HQET operators
with the appropriate quantum numbers and dimensionalities.
At the leading (zeroth) order in $1/m$,
we encounter the HQET heavy-light current:
\begin{equation*}
\widetilde{\jmath}(\mu) = \widetilde{Z}_j^{-1}(\alpha_{\text{s}}(\mu)) \widetilde{\jmath}_0\,,
\quad
\widetilde{\jmath}_0 = \bar{q}_0 \Gamma \widetilde{Q}_0\,,
\end{equation*}
where $\Gamma$ is a Dirac matrix.

Let the sum of one-particle-irreducible bare diagrams
with the current vertex $\Gamma$,
an incoming quark $Q$ with residual energy $\omega$,
and an outgoing quark $q$ with momentum $p$,
not including the external quark propagators,
be the proper vertex $\widetilde{\Gamma}(\omega,p)=\Gamma+\widetilde{\Lambda}(\omega,p)$
(here $\Gamma$ in the right-hand side is the Dirac matrix).
When it is expressed via the renormalized quantities
$\alpha_{\text{s}}(\mu)$, $a(\mu)$,
it should become $\widetilde{Z}_\Gamma\widetilde{\Gamma}_{\text{r}}(\omega,p)$,
where $\widetilde{Z}_\Gamma$ is minumal~(\ref{minim}),
and the renormalized vertex $\widetilde{\Gamma}_{\text{r}}(\omega,p)$
is finite in the limit $\varepsilon\to0$.
When the proper vertex of the renormalized current
$\widetilde{Z}_j^{-1}\widetilde{\Gamma}(\omega,p)$
is multiplied by the two external leg renormalization factors $Z_q^{1/2}\widetilde{Z}_Q^{1/2}$,
it should give a finite matrix element.
Therefore, $\widetilde{Z}_j=\left(Z_q\widetilde{Z}_Q\right)^{1/2}\widetilde{Z}_\Gamma$.

\begin{figure}[ht]
\begin{center}
\begin{picture}(32,9.5)
\put(16,4.75){\makebox(0,0){\includegraphics{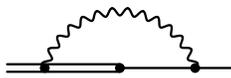}}}
\end{picture}
\end{center}
\caption{One-loop vertex}
\label{F:V1}
\end{figure}

Ultraviolet divergences of $\widetilde{\Lambda}(\omega,p)$
don't depend on quark masses and the external momenta.
Therefore, we may assume that all quarks are massless,
and set $\omega=0$, $p=0$.
An infrared cutoff is then necessary in order to avoid
infrared $1/\varepsilon$ terms.
In the one-loop approximation (Fig.~\ref{F:V1}),
\begin{equation*}
\begin{split}
\widetilde{\Lambda}(0,0) &{}= - \I C_{\text{F}} g_0^2 \int \frac{\D^d k}{(2\pi)^d}
\frac{\gamma^\mu \rlap/k v^\nu
\left[g_{\mu\nu} - (1-a_0) k_\mu k_\nu / k^2\right]}{(k^2)^2 k_0}\\
&{}= - \I C_{\text{F}} g_0^2 \int \frac{\D^d k}{(2\pi)^d}
\frac{\gamma_0\rlap/k - (1-a_0) k_0}{(k^2)^2 k_0}\,.
\end{split}
\end{equation*}
Here $\rlap/k=k_0\gamma_0-\vec{k}\cdot\boldsymbol{\gamma}$;
the term with $\vec{k}$ yields 0 after integration:
\begin{equation*}
\widetilde{\Lambda}(0,0) = - \I C_{\text{F}} g_0^2 a_0
\int \frac{\D^d k}{(2\pi)^d} \frac{1}{(k^2)^2}\,.
\end{equation*}
The ultraviolet $1/\varepsilon$ pole of this integral is
\begin{equation*}
\begin{split}
\left. \int \frac{\D^d k}{(2\pi)^d}\; \frac{1}{(k^2)^2} \right|_{\text{UV}}
&{} = \I \frac{2\pi^{d/2}}{(2\pi)^d\Gamma(d/2)} \int_\lambda^\infty k^{-1-2\varepsilon} \D k\\
&{} = - \frac{\I}{(4\pi)^2 \varepsilon} \left.k^{-2\varepsilon}\right|_\lambda^\infty
= \frac{\I}{(4\pi)^2 \varepsilon}\,,
\end{split}
\end{equation*}
where $\I$ comes from the rotation to the Euclidean space,
$\lambda$ is the infrared cutoff,
and terms regular at $\varepsilon\to0$ are omitted.
Therefore,
\begin{equation*}
\widetilde{Z}_\Gamma = 1 + a C_{\text{F}} \frac{\alpha_{\text{s}}}{4\pi\varepsilon}\,.
\end{equation*}
Taking into account $\widetilde{Z}_Q$~(\ref{ZQ1}) and
\begin{equation*}
Z_q = 1 - a C_{\text{F}} \frac{\alpha_{\text{s}}}{4\pi\varepsilon}\,,
\end{equation*}
we see that the gauge dependence cancels:
\begin{equation*}
\widetilde{Z}_j = 1 + \frac{3}{2} C_{\text{F}} \frac{\alpha_{\text{s}}}{4\pi\varepsilon}\,.
\end{equation*}
Finally,
\begin{equation*}
\widetilde{\gamma}_j = - 3 C_{\text{F}} \frac{\alpha_{\text{s}}}{4\pi}\,.
\end{equation*}

\begin{figure}[ht]
\begin{center}
\begin{picture}(106,60.5)
\put(53,30.25){\makebox(0,0){\includegraphics{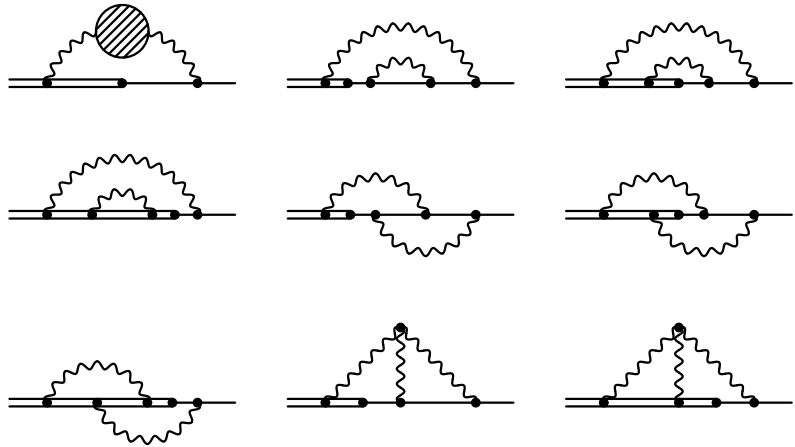}}}
\end{picture}
\end{center}
\caption{Two-loop vertex}
\label{F:V2}
\end{figure}

We could equally easily obtain this result by calculating
$\widetilde{\Lambda}(\omega,0)$ without infrared cutoff.
Keeping $\omega\ne0$ is enough to ensure infrared convergence.
At two loops this becomes mandatory.
Calculating $\widetilde{\Lambda}(\omega,0)$ in the two-loop approximation
(Fig.~\ref{F:V2}) by the methods of Sect.~\ref{S:2l},
we can obtain the anomalous dimension~\cite{BG:91}
\begin{align}
\widetilde{\gamma}_j ={}& - 3 C_{\text{F}} \frac{\alpha_{\text{s}}}{4\pi}
\label{gamma2}\\
&{} + C_{\text{F}} \left[ \left(-\frac{8}{3}\pi^2+\frac{5}{2}\right) C_{\text{F}}
+ \left(\frac{2}{3}\pi^2-\frac{49}{6}\right) C_{\text{A}}
+ \frac{10}{3} T_{\text{F}} n_{\text{f}}
\right] \left(\frac{\alpha_{\text{s}}}{4\pi}\right)^2 + \cdots
\nonumber
\end{align}

\end{document}